# Thermal conductivity of disordered porous membranes

Marianna Sledzinska[1,*], Bartlomiej Graczykowski[2,3], Francesc Alzina[1], Umberto Melia[4], Konstantinos Termentzidis[5], David Lacroix[6] and Clivia M. Sotomayor Torres[1,7]

[1] Catalan Institute of Nanoscience and Nanotechnology (ICN2), CSIC and The Barcelona Institute of Science and Technology Campus UAB, Bellaterra, 08193 Barcelona, Spain

[2] NanoBioMedical Centre, Adam Mickiewicz University, ul. Umultowska 85, PL-61614 Poznan, Poland

[3] Max Planck Institute for Polymer Research, Ackermannweg 10, 55218 Mainz, Germany

[4] Department of ESAII, Centre for Biomedical Engineering Research, Universitat Politècnica de Catalunya, CIBER-BBN, Barcelona, Spain

[5] Univ Lyon, CNRS, INSA-Lyon, Université Claude Bernard Lyon 1, CETHIL UMR5008, F-69621, Villeurbanne, France

[6] Université de Lorraine, CNRS, LEMTA, Nancy, F-54000, France

[7] ICREA - Institucio Catalana de Recerca i Estudis Avancats, 08010 Barcelona, Spain

*corresponding author marianna.sledzinska@icn2.cat

We report measurements and Monte Carlo simulations of thermal conductivity of porous 100nm- thick silicon membranes, in which size, shape and position of the pores were varied randomly. Measurements using 2-laser Raman thermometry on both plain membrane and porous membranes revealed more than 10-fold reduction of thermal conductivity compared to bulk silicon and six-fold reduction compared to non-patterned membrane for the sample with 37% filling fraction.

Using Monte Carlo solution of the Boltzmann transport equation for phonons we compared different possibilities of pore organization and its influence on the thermal conductivity of the samples. The simulations confirmed that the strongest reduction of thermal conductivity is achieved for a distribution of pores with arbitrary shapes that partly overlap. Up to 15% reduction of thermal conductivity with respect to the purely circular pores was predicted for a porous membrane with 37% filling fraction. The effect of pore shape, distribution and surface roughness is further discussed.

**Introduction**

Although physical models like to describe perfectly ordered systems, they are hard to find in nature. The real systems always account for the presence of defects and disorder. In many systems the presence of disorder is considered unfavorable, however recent works have shown that it can improve or add functionality to the system. For instance, in nature disordered nanostructures enable flowers to produce visual signals that are salient to bees [1]. In photonics, photon transport and collimation can be strongly enhanced in disordered structures [2].

Thermal transport in a porous material is influenced by several factors such as porosity, pore size and shape, temperature and emissivity in the pore [3]. At the same time disorder in the crystalline lattice, such as presence of grain boundaries, defects and heavy doping reduces thermal conductivity of the material [4]. We have previously investigated the influence of short-range disorder in Si membrane-based two-dimensional phononic crystals on the GHz and THz phononic properties. We showed that although low-frequency phonon modes are affected by periodicity, their impact is not sufficient to affect thermal conductivity at room temperature. For samples with period of 300 nm we observed purely diffusive thermal transport and obtained the same value for the thermal conductivity for the ordered and disordered PnCs [5].

Indeed, when the pore and pitch sizes are in the order of few hundreds of nanometers the position of each pore has been reported to play a crucial role mainly at very low temperatures. Zen et al. showed coherent reduction of thermal conductance at sub-kelvin temperatures [6] and Maire et al. demonstrated the transition from coherent to purely diffusive heat conduction at 10 K [7]. In a paper by Anufriev et al. the effect of disorder in circular pores of PnCs was further studied. Indeed in that study reduction of thermal conductivity at room temperature induced by disorder was only measurable in the samples with high filling fraction and periodicity below 200nm [8].

In this work we did not only change the position of the holes and filling fraction, we also randomized the size and shape of the hole. The detailed sample characterization was performed using scanning electron microscopy (SEM). The thermal conductivity of the samples was measured using 2-laser Raman thermometry (2LRT) and the results were compared with Monte Carlo simulations of phonon transport in similar membranes.

**Sample description**

Disordered porous membranes can be fabricated using variety of techniques, such as nanocrystallisation, block-copolymer lithography or porous alumina templates. Here in order to fabricate the disordered structures we have modified the fabrication process based on electron beam lithography (EBL) and reactive ion etching (RIE) on free-standing membranes (Norcada Inc.) [9]. EBL systems are designed to be robust to external noise in order to reproduce the design with the highest accuracy and the influence of shot noise for example is reduced [10]. To intentionally introduce disorder in our structures, we recurred to a noise-assisted EBL, where strong external vibrations distort the beam path. In this case, the targeted design of a periodic porous membrane, similar to the ones reported previously [11], becomes a randomized lattice of deformed holes.

The two structures, hereinafter referred as Sample 1 and Sample 2, fabricated in a membrane with thickness $t = 100$ nm are shown in Fig 1 (a) and (c), respectively. The samples were carefully imaged using SEM and images were further analyzed using Matlab software (see Appendix A). First of all, the effective filling fraction was calculated for Samples 1 and 2 to be of 25 and 37%, respectively. The shapes of the pores are not circular, therefore we first calculate the distribution of the areas of the pores. Due to the computational reasons, in the MC model we assume circular distribution of the pores, however maintaining the filling fraction and the distribution of the areas. The comparison between cumulative sum of the areas of the pores obtained experimentally and used in the simulations is shown in Fig. 1 (b) and (d). If we assume the circular shape of the pores the corresponding distribution of the diameters is shown in the insets in Fig.1.

**Experimental data**

Thermal conductivity measurements were performed using 2LRT technique which has been successfully applied to 2D free-standing plain [12, 13] and porous membranes [5, 11]. The technique consists of using two lasers, one of which (405nm wavelength) generates a localized steady-state thermal excitation, while the second laser (532nm wavelength) measures the spatially resolved temperature profile through the temperature-dependent Raman frequency of the optical phonons in the material [12]. Both lasers were focused on the samples using 50× microscope objectives with numerical apertures of NA = 0.55. Both samples have a non-patterned 5x5 µm² square in the center (cf. Fig 1 (c)) where the heating laser is focused. Finally the absorbed power $P_{abs}$ is measured for each sample as the difference between incident and transmitted plus reflected light intensities probed by a calibrated system based on a non-polarizing cube beam splitter. All the measurements were performed in a vacuum of 5x10$^{-3}$ mbar.

Solving Fourier's law in 2D for a thermally isotropic medium in steady state we obtain:

$$\frac{P_{abs}}{2\pi rt} = -k\frac{dT}{dr}$$

where $r$ is the distance from the heating spot, $t$ is the membrane thickness, $k$ is the thermal conductivity and $T$ is the temperature of the sample.

First of all the, non-patterned silicon membrane with $t$ = 100nm was measured as a reference sample. The temperature profile for an absorbed power $P_{abs}$=2.18 mW is shown in Fig. 2(a). When the variation of $k$ is small in the temperature range under study, it can be extracted from Eq. (1) and the slope of the temperature profile as a function of ln(r). The solid line in Fig 2(d) is the linear fit of the experimental data (black circles) from which a thermal conductivity $k$ of 60 +/- 3 Wm$^{-1}$K$^{-1}$ was obtained. This represents more than a two-fold reduction compared to bulk silicon [14].

Secondly, the two porous samples were measured. The temperature profiles for Samples 1 and 2 are shown in Fig. 2 (b) and (c), respectively, for a given $P_{abs}$. The extracted values of the intrinsic thermal conductivity for these samples are 19 +/- 3 Wm$^{-1}$K$^{-1}$ and 11 +/-3 Wm$^{-1}$K$^{-1}$ for the Sample 1 and 2, respectively (Fig 2 (e),(f)). This means that thermal conductivity in the disordered porous membrane configuration is lowered by one order of magnitude as compared to bulk.

**Monte Carlo simulations**

Boltzman transport equation (BTE) can be handled with different techniques, particularly those developed around time and space discretization (finite volumes, finite elements, etc.) and which are efficient for simple geometries. In the present case, membranes with disordered distribution of pore size and location, meshing such nano-device is challenging and make finite difference schemes very complex to use. In such complex geometrical configuration, Monte Carlo technique based on energy carrier displacement and scattering is much more appropriate and has already proven to be very efficient and reliable. The MC solution of BTE used in this study is derived from a previously developed simulation tools that allows the modelling of plain membranes [15] and porous membranes [16]. In the latter works mono-dispersed spherical pores or aligned/staggered cylindrical pores where either considered. The main evolution of the MC tool in the present work was to make possible the modelling of poly-dispersed cylindrical pores that can overlap which mimic the disordered porous membranes experimentally achieved (Fig. 3). All the basics of phonon transport in nanostructures (drift and scattering of carriers) with MC method can be found in the previously cited papers. Concerning physical inputs, the dispersion properties of bulk silicon, assuming isotropic dispersion relations, are used. Similarly, phonon lifetimes are derived from bulk according to the Holland formalism [17]. Same as in the experiment, membrane thickness is taken to be 100nm and random distribution of holes for a porosity of 25% and 37% with no overlapping of holes was studied. Diffuse reflections are assumed from pore boundaries and membrane's surfaces. The width of the samples was varied from 10 to 40 µm and length of the porous part was set to 2µm, as shown in Fig. 3. The temperatures set at the hot and cold end were 302K and 298K, respectively. In order to reduce the uncertainty the results were averaged over 16 different membrane configurations for each porosity and overlapping parameters.

First of all, thermal conductivity of a plain 100 nm thick membrane was calculated to be 59.26 +/-1.33 $Wm^{-1}K^{-1}$, in good agreement with the value obtained experimentally. Then we proceed to calculate the thermal conductivity of the porous membranes, assuming circular shape of the pores and same filling fraction. The results obtained for the samples 1 and 2 are of 35 and 30 $Wm^{-1}K^{-1}$, respectively. We will discuss these results in detail in the following part.

**Discussion**

To the best of our knowledge, we have only found one paper dealing with disordered porous membranes with similar geometric parameters. In the work by Wolf et al. numerical simulations were carried out on porous thin films with thickness of 100 nm and pore diameter of 50 nm, different pore roughness, from diffuse to specular, and random pore locations in the nanostructure [18]. Among the different outputs of this work three main conclusions were drawn. First, increasing porosity leads to thermal conductivity reduction: a three-fold reduction was calculated from unpatterned to 35% of porosity. This has been observed in several studies dealing with nanoporous materials, and the results are in the same range as the results obtained in our work. Second, increasing pore roughness, i.e. assuming that pores diffusely scatter phonons, also reduces the thermal conductivity; this is also observed in our simulation with a limited impact as porosity increases. Once again, this is expected as large porosity makes the overall behavior of the structure diffuse regarding the phonon transport and increasing this diffuse behavior by pore scattering do not affect the calculations. Lastly, the impact of pore placement in the membrane, meaning either a uniform or a non-uniform distribution on the resulting thermal conductivity was studied. This is undoubtedly the most complicated issue to deal with, as it can change drastically the value of the thermal conductivity up to an order of magnitude depending on the pore placement. Two extreme cases were applied: the pore can be uniformly located and thermal transport is only ruled by porosity and specularity at pore interface, or pores can be non-uniformly placed and locally increase the resistance to phonon transport due to the fact that some areas of the membrane have increased apparent porosity.

The same effect has also been recently discussed by Romano et al., who emphasized the role of special pore arrangements, so called phonon bottlenecks [19]. A phonon bottleneck is a set of pores representing the highest local resistance to phonon transport. Even though the simulations were performed for much smaller structures (period 10 nm) the results are similar to the ones of Wolf et al.. The formation of heat high-flux regions is irregular as it depends on the pore configuration and the density of pores along the heat flux direction has a significant influence on thermal conductivity. The disordered circular crystals were found to have average thermal conductivity $k$ values up to 15% lower than that of their aligned counterparts. This is in line with what was observed in our samples, where some parts of the surface exhibit higher pore density than other (see Fig 1). This naturally affects the resulting $k$ both in experimental and MC simulations.

Apart from the particular pore placement it was previously reported that different shapes of pores (circular, square and triangular) would give varying thermal conductivity, however these simulations were made for pores of significantly smaller size of about few nanometers [20]. In case of disordered lattice of square pores the $k$ values were 30% lower than that of their aligned counterparts [19]. Therefore, there from MC simulations reported previously there is a clear connection between the position and the shape of the pores and the thermal conductivity.

In fact, the main difference between the experimental samples and samples developed for simulations is the shape of the pores. While in the model we assume circular shape for computational simplicity, the actual shape of the pores in the fabricated sample is arbitrary. In order to partly reproduce this randomness in shape, holes are allowed overlapping, while always maintaining the filling fraction constant (Fig. 3). Overlapping is introduced by weakening the limitation introduced in the previous simulations that the minimal distance between pores is when they edges are touching, i.e., $R_i + R_j$ with $R_i$ and $R_j$ the pore radii. This distance is now established to vary between 40 and 85% of the previously set value $R_i + R_j$. The results for different percentage of amount of overlapping are summarized in Table 1. For sample 1 there is a slight decrease of about 2 Wm$^{-1}$K$^{-1}$ in thermal conductivity for a 55% of overlapping. More significant results are obtained for ample 2. In this case we obtained a decrease of 4.5 Wm$^{-1}$K$^{-1}$ for 55% overlapping. In both simulated samples, $k$ is seen to reach a minimal value at 55 % of overlapping, and increasing with a further decrease of the center to center distance. This seems to be related to the optimal value for overlapping giving maximum blocking of the direct paths of heat flow.

However, even for the overlapping pores there is still the discrepancy between experimental and Monte Carlo results of about 10 Wm$^{-1}$K$^{-1}$. Apart from the difference in shape of the pores, below we discuss other possible reasons.

One effect, which accounts for the low $k$ of experimental samples, is the roughness at the edges of the pores. We have estimated that the roughness for samples fabricated using EBL and RIE to be below 5nm [5]. However it is more pronounced in the samples reported in this work, approximately of 7 nm. This roughness comes from the particular fabrication process applied here. It would be logic to assume that this layer inside the pores has lower thermal conductivity that the crystalline silicon of the membrane. As Verdier et al. have showed that amorphous layers on the pores' surface can reduce the overall thermal conductivity of the system [21]. They affect thermal transport through possible localized modes that do not contribute to the energy transport in the material. Therefore we believe that the important amount of the partly amorphised layer adds to overall low thermal conductivity.

Finally, from the SEM images we also note that there is a non-negligible amount of small pores (approx. 50nm diameter) which are not fully etched through (Fig. A1.). They were not taken into account while calculating the filling fraction, but they contribute to the partial reduction of the thermal conductivity, as their phonon scattering efficiency is larger than big pores. All this phenomena might lower thermal conductivity, which was checked numerically carrying out MC simulation with pore size distribution including pores as small as 40nm in diameter.

| Sample 1 | | Sample 2 | | |
|---|---|---|---|---|
| $k$ (Wm$^{-1}$K$^{-1}$) | Porosity (%) | $k$ (Wm$^{-1}$K$^{-1}$) | Porosity (%) | Overlapping (%) |
| 59.26 +/- 1.33 | 0 | 59.26 +/- 1.33 | 0 | - |
| 35.61 +/- 0.14 | 25 | 30.01 +/- 0.14 | 37 | No |
| 33.12 +/- 0.53 | 26.5 | 27.41 +/- 0.14 | 38.1 | 85 |
| 32.89 +/- 0.14 | 25.5 | 27.45 +/- 0.15 | 38 | 70 |
| 32.86 +/- 0.13 | 25.5 | 25.66 +/- 0.56 | 38 | 55 |
| 33.39 +/- 0.13 | 25.1 | 26.97 +/- 0.56 | 36.5 | 40 |
| 19 +/- 3 (experiment) | 25 | 11 +/-3 (experiment) | 37 | - |

Table 1 Summary of the Monte Carlo and experimental results for the cases of varying possible overlap percentage of the pores, resembling sample 1 and sample 2.

**Conclusions**

In this work we have shown that by introducing disorder not only in the location of the pores, but also in the shape we can effectively reduce thermal conductivity. This effect is more pronounced for samples with high filling fraction. According to MC simulations, for the sample with 37% filling fraction a decrease in thermal conductivity of approximately 15% can be achieved by introducing 55% of overlap between the pores. This could have an important impact of fabrication of porous membranes for thermoelectric applications: in order to reduce thermal conductivity it is not necessary to increase the porosity of the samples, as the same effect can be achieved by varying the position and shape of the pores. More systematic experimental studies are needed to evaluate these effects, especially considering the length scales in which these effects are playing an important role. We have also discussed other possible factors which can contribute to the decrease of thermal conductivity, such as roughness and presence of not fully etched-through pores.

**Acknowledgments**

The ICN2 is funded by the CERCA programme / Generalitat de Catalunya. The ICN2 is supported by the Severo Ochoa program of the Spanish Ministry of Economy, Industry and Competitiveness (MINECO, grant no. SEV-2013-0295). We acknowledge the financial support from the Spanish MINECO project PHENTOM (FIS2015-70862-P). We acknowledge the financial support from the French ANR with the project MESOPHON (ANR-15-CE30-0019).

Appendix A: Porosity analysis

Grayscale images obtained from the SEM (Fig. A.1 (a)) were converted in binary format by applying a threshold on the pixel values. For each image, the respective threshold was obtained by visual analysis of the image intensity histogram. Figure A.1 shows an example of an image histogram and the respective threshold. The importance of choosing the correct threshold is shown in fig A1. In the (c) and (d) we compare two images obtained with setting the threshold at 150 and 200, respectively. The feature marked with red circle shows three pores: one etched completely and two etched half-through. If the threshold is set too low (at 150 in this case) all three pores appear in the binary image. If the threshold is set correctly (at 200 in this case) only the fully etched pore appear in the binary image.

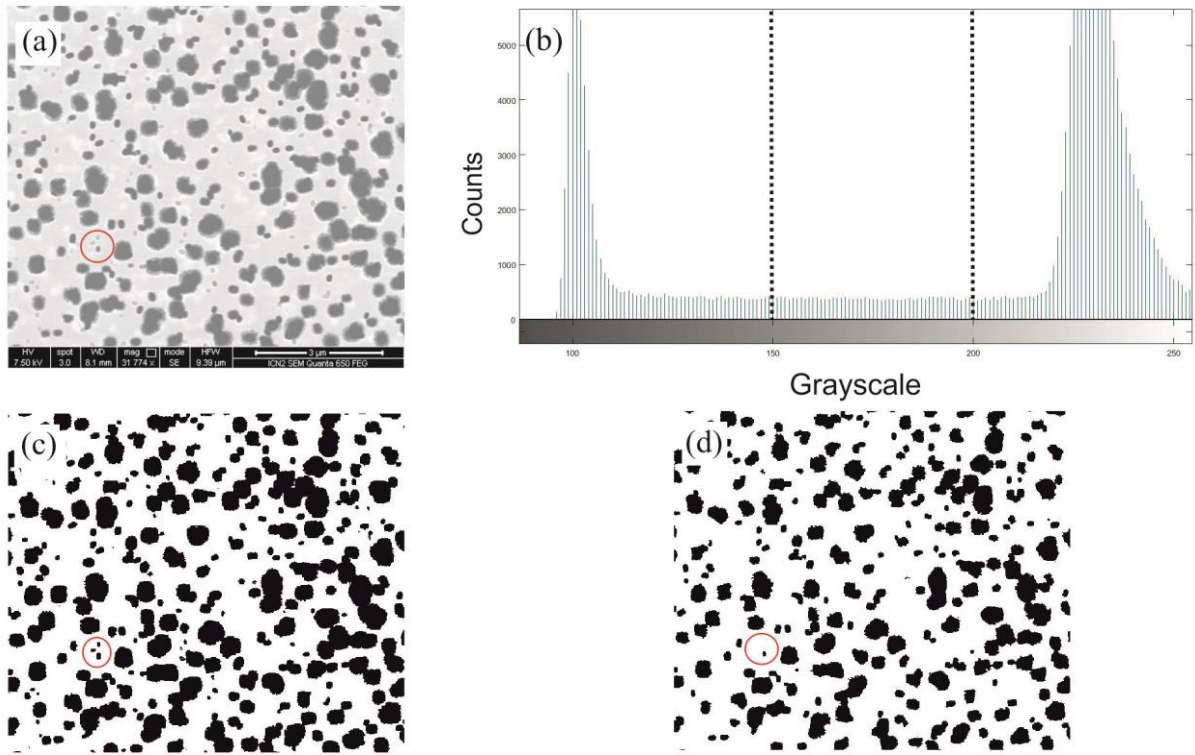

Figure A.1 (a) Original SEM image in grayscale, scale bar 3µm; (b) Image histogram. Image in binary format after the application of the threshold at (c) 150 (d) 200.

The area of each pore was calculated by using the Matlab function "*regionprops*" that returns the amount of pixel included in each black region of the image. The area in pixel was converted in nm² by using the scale indicated at the bottom of the figure and the following equation:

$$Area(nm^2) = Area(pixel) \frac{scale(nm)^2}{(X_{s1} - X_{s2})^2}$$

Where $X_{s1}$ and $X_{s2}$ are the x coordinates of the scale bar extremities in pixel unit.

Finally, a histogram that shows the distribution of the areas of all the pores was obtained (as in figure 1). Area values below 1260 nm² were discarded as they were considered as pixel noise.

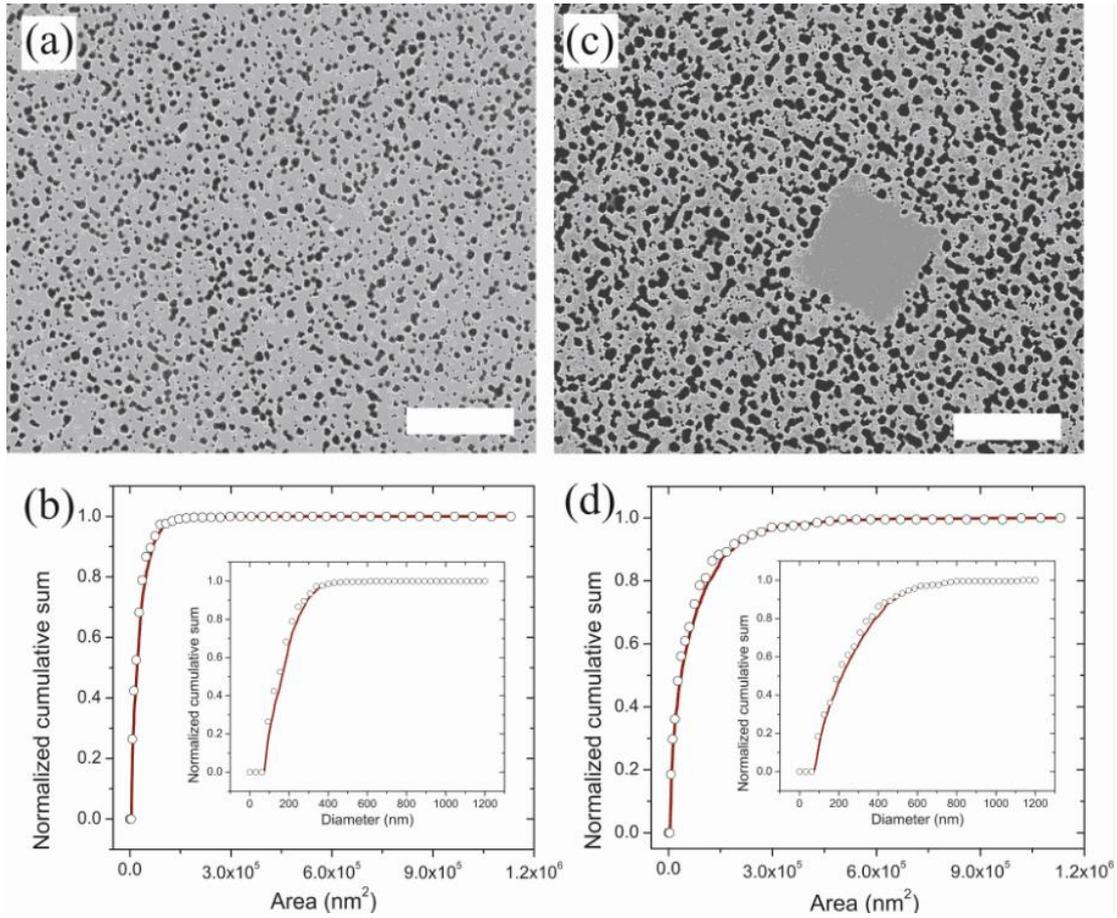
Figure 1 SEM images of (a) Sample 1 with 25% filling fraction. (c) Sample 2 with 37% filling fraction and the detail of heating island located in the center of the sample. Scale bar 5um. Normalized cumulative sum of the areas of the pores in (b) Samples 1, (d) Sample 2. Insets correspond to the cumulative sum of the diameters, if we assume circular pores for a given area distribution.

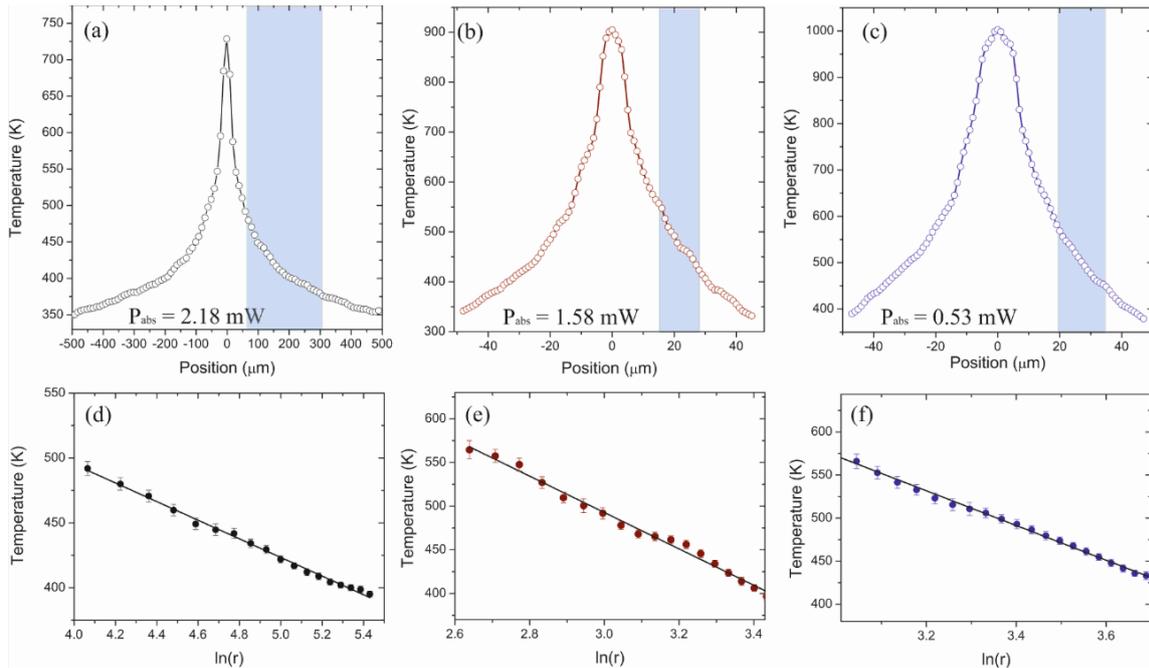

Figure 2 Temperature profiles and temperatures as a function of ln(r) for: (a), (d) non-patterned, 100nm thick membrane. (b), (e) Disordered porous membranes with filling fraction of 25%. (c), (f) Disordered porous membranes with filling fraction of 37%

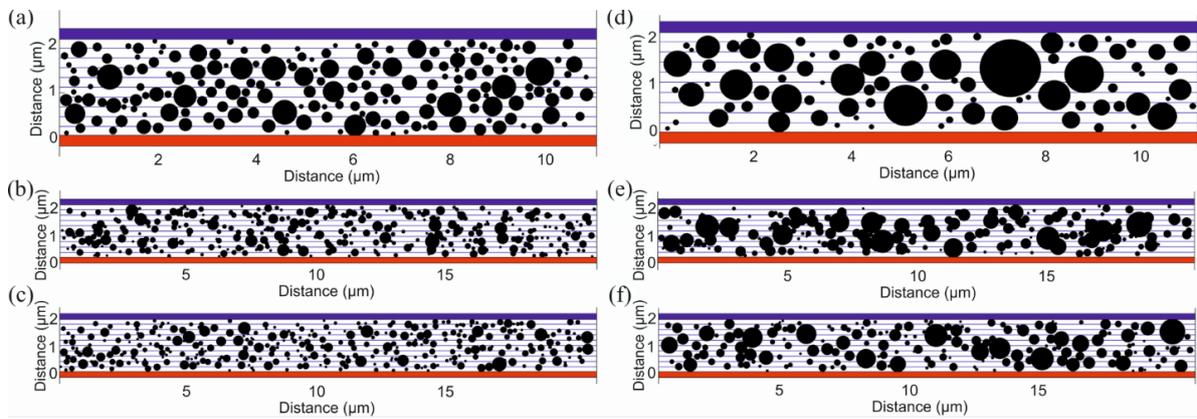

Figure 3 Pore distribution used in Monte Carlo simulations. Sample 1: (a) No overlap, (b) 55% possible overlap (c) 85% possible overlap. Sample 2: (d) No overlap, (e) 55% possible overlap (f) 85% possible overlap.